\documentclass[printer]{aa}
\usepackage{psfig,txfonts}

\def\ltsima{$\; \buildrel < \over \sim \;$}
\def\lsim{\lower.5ex\hbox{\ltsima}}
\def\gtsima{$\; \buildrel > \over \sim \;$}
\def\gsim{\lower.5ex\hbox{\gtsima}}

\begin{document}

\title{Late-epoch optical and near-infrared observations \\ of the
GRB000911 afterglow and its host galaxy\thanks{Based on observations
made with ESO Telescopes at the Paranal and La Silla Observatories under
programmes 165.H-0464 and 265.D-5742, with the Italian Telescopio
Nazionale Galileo and with the Nordic Optical Telescope (NOT), both 
operating on the island of La Palma in the Spanish Observatorio del Roque 
de los Muchachos of the Instituto de Astrof\'{i}sica de Canarias. The NOT 
is operated jointly by Denmark, Finland, Iceland, Norway, and Sweden.}}

\titlerunning{Late-epoch optical/NIR study of GRB000911 and its host}
\authorrunning{N. Masetti et al.}

\author{N. Masetti\inst{1}, 
E. Palazzi\inst{1}, 
E. Pian\inst{1,2},
L. Hunt\inst{3},
J.P.U. Fynbo\inst{4},
J. Gorosabel\inst{5},
S. Klose\inst{6},
S. Benetti\inst{7}, \\
R. Falomo\inst{7}, 
A. Zeh\inst{6},
L. Amati\inst{1},
M.I. Andersen\inst{8},
A.J. Castro-Tirado\inst{5},
J.M. Castro Cer\'on\inst{4,9}, \\
J. Danziger\inst{2},
F. Frontera\inst{1,10},
A.S. Fruchter\inst{9},
J. Greiner\inst{11},
J. Hjorth\inst{4},
B.L. Jensen\inst{4},
L. Kaper\inst{12}, \\
C. Kouveliotou\inst{13},
A. Levan\inst{9,14},
A. Magazz\`u\inst{15},
P. M\o ller\inst{16},
L. Nicastro\inst{17},
H. Pedersen\inst{4},
N.R. Tanvir\inst{18}, \\
P.M. Vreeswijk\inst{19},
R.A.M.J. Wijers\inst{12} and
E.P.J. van den Heuvel\inst{12}
}

\institute{
INAF - Istituto di Astrofisica Spaziale e Fisica Cosmica, Sezione di Bologna, 
Via Gobetti 101, I-40129 Bologna, Italy (formerly IASF/CNR, Bologna)
\and
INAF - Osservatorio Astronomico di Trieste, Via G.B. Tiepolo 11, I-34131
Trieste, Italy
\and
INAF - Istituto di Radioastronomia, Sezione di Firenze, largo E. 
Fermi 5, I-50125, Florence, Italy
\and
Niels Bohr Institute, University of Copenhagen, Juliane Maries Vej 30,
DK-2100 Copenhagen {\O}, Denmark
\and
Instituto de Astrof\'{\i}sica de Andaluc\'{\i}a (IAA-CSIC),
Apartado de Correos 03004, E-18080 Granada, Spain
\and
Th\"uringer Landessternwarte Tautenburg, D-07778 Tautenburg, Germany
\and
INAF - Osservatorio Astronomico di Padova, vicolo dell'Osservatorio 5,
I-35122 Padua, Italy
\and
Astrophysikalisches Institut Potsdam, An der Sternwarte 16, D-14482 
Potsdam, Germany
\and
Space Telescope Science Institute, 3700 San Martin Drive, Baltimore, MD
21218, USA
\and
Dipartimento di Fisica, Universit\`a di Ferrara, via Paradiso 12, I-44100
Ferrara, Italy
\and
Max-Planck-Institut f\"ur Extraterrestrische Physik, 
Giessenbachstrasse, D-85748 Garching, Germany
\and
Institute of Astronomy ``Anton Pannekoek", University of Amsterdam,
Kruislaan 403, 1098 SJ Amsterdam, The Netherlands
\and
NASA Marshall Space Flight Center, SD-50, Huntsville, AL 35812, USA
\and
Department of Physics and Astronomy, University of Leicester, University
Road, Leicester, LE1 7RH, United Kingdom
\and
INAF - Telescopio Nazionale Galileo, Roque de Los Muchachos Astronomical
Observatory, P.O. Box 565, E-38700 Santa Cruz de La Palma, Spain
\and
European Southern Observatory, Karl Schwarzschild-Strasse 2, D-85748 
Garching, Germany
\and
INAF - Istituto di Astrofisica Spaziale e Fisica Cosmica, Sezione di Palermo, 
via La Malfa 153, I-90146 Palermo, Italy (formerly IASF/CNR, Palermo)
\and
Centre for Astrophysics Research, University of Hertfordshire,
College Lane, Hatfield, Herts AL10 9AB, UK
\and
European Southern Observatory, Casilla 19001, Santiago 19, Chile
}

\offprints{N. Masetti, (\texttt{masetti@bo.iasf.cnr.it})}
\date{Received 25 January 2005; accepted 18 April 2005}

\abstract{
We present the results of an optical and near-infrared (NIR) monitoring
campaign of the counterpart of Gamma-Ray Burst (GRB) 000911, located at
redshift $z$ = 1.06, from 5 days to more than 13 months after explosion.
Our extensive dataset is a factor of 2 larger and spans a time
interval $\sim$4 times longer than the ones considered previously for this
GRB afterglow; this allows a more thorough analysis of its light curve and
of the GRB host galaxy properties. The afterglow light curves show a
single power-law temporal decline, modified at late times by light from a
host galaxy with moderate intrinsic extinction, and possibly by an
emerging supernova (SN). The afterglow evolution is interpreted within the
classical ``fireball" scenario as a weakly collimated adiabatic shock
propagating in the interstellar medium. The presence of a SN light curve
superimposed on the non-thermal afterglow emission is investigated: while
in the optical bands no significant contribution to the total light is
found from a SN, the NIR $J$-band data show an excess which is consistent
with a SN as bright as the known hypernova SN1998bw. If the SN
interpretation is true, this would be the farthest GRB-associated SN, as
well as the farthest core-collapse SN, discovered to date. However, other
possible explanations of this NIR excess are also
investigated. Finally, we studied the photometric properties of the
host, and found that it is likely to be a slightly reddened,
subluminous, extreme starburst compact galaxy, with luminosity
$\sim$0.1$L^\star$, an age of $\sim$0.5 Gyr and a specific Star Formation
Rate (SFR) of $\approx$30 $M_\odot$ yr$^{-1}$ ($L/L^\star$)$^{-1}$. This
is the highest specific SFR value for a GRB host inferred from optical/NIR
data.

\keywords{gamma rays: bursts --- supernovae: general --- radiation 
mechanisms: non-thermal --- cosmology: observations --- galaxies: 
high-redshift --- galaxies: fundamental parameters}}

\maketitle

\section{Introduction}

Long and frequent ground-based monitoring at optical and near-infrared 
(NIR) wavelengths of Gamma-Ray Burst (GRB) counterparts is necessary to
constrain the emission models and disentangle the contribution of the
GRB host galaxy and of a possible underlying supernova (SN) explosion 
simultaneous with or shortly preceding the GRB (e.g., McFadyen \& Woosley 
1999; Vietri \& Stella 1998).

The connection of GRBs with SNe was first suggested by the close angular
and temporal proximity of GRB980425 and SN1998bw (Galama et al. 1998) and,
based on that event, this association was systematically explored (e.g.,
Castro-Tirado \& Gorosabel 1999; Bloom et al. 1999: Galama et al. 2000;
Reichart 2001; Garnavich et al. 2003; Greiner at al. 2003; Price et al.
2003a; Masetti et al. 2003). The recent direct spectroscopic detections of
SNe 2002lt (Della Valle et al. 2003), 2003dh (Stanek et al. 2003; Hjorth et
al. 2003) and 2003lw (Malesani et al. 2004) associated with GRBs 021211,
030329 and 031203, respectively, firmly established that at least a
fraction of GRBs is connected with core-collapse SNe.

However, given the high redshifts ($z \ga 1$) of most GRBs, SNe associated
with them are difficult to study spectroscopically, and only good
observational coverage of the optical light curve can reveal the presence
of an underlying SN (see Zeh et al. 2004; Dar \& De R\'ujula 2003). In
addition, due to the possible dust obscuration near the source, NIR data
may play an even more important role than optical data (e.g., Palazzi et
al. 1998; Gorosabel et al. 1998; Pian et al. 1998; Vreeswijk et al. 1999;
Castro-Tirado et al. 1999; Masetti et al. 2000; Klose et al. 2000; Rhoads
\& Fruchter 2001; Sokolov 2001; Le Floc'h et al. 2003). Thus, indirect
(i.e. late-epoch photometric) observations of the presence of a SN are
still important in the study of the GRB-SN connection.

GRB000911 was detected by the Inter-Planetary Network (IPN) satellites on
2000 September 11.30237 UT as a very long ($\sim$500 s) burst (Hurley et
al.  2000; Price et al. 2002). Shortly thereafter, Berger et al. (2000)
found a variable radio source inside the GRB error box and, at the radio
position, Price et al. (2000, 2002) detected an optical transient (OT) at
coordinates RA = 02$^{\rm h}$ 18$^{\rm m}$ 34$\fs$36, Dec = +07$^{\circ}$
44$'$ 27$\farcs$65 (J2000) which faded according to a power-law decay
$F$($t$)  $\propto t^{-\alpha}$ with index $\alpha$ = 1.46. This
behaviour, together with the positional coincidence with the mentioned
transient radio source, strongly suggested that this OT was indeed the
optical afterglow of GRB000911.

Smith et al. (2001) reported no detection of transient emission at 
sub-mm (850 $\mu$m) wavelengths.
In this same band, Berger et al. (2003) reported the detection of emission 
from the host galaxy of GRB000911; this detection was however questioned by 
Tanvir et al. (2004) on the basis of the former sub-mm upper limit of Smith 
et al. (2001) for the combined host plus afterglow emission.
Extensive optical and NIR observations of the afterglow have been reported
by Lazzati et al. (2001), who inferred the presence of a SN reaching
maximum light at about 18 days after the GRB (in the rest frame at $z$ =
1.0585: Djorgovski et al. 2001; Price et al. 2002). The underlying 
host galaxy appears to be a starburst with moderate dust absorption, i.e. 
with 0.11 $< E(B-V) <$ 0.21 (Lazzati et al. 2001). No follow-up 
observations of this GRB were performed in X--rays.

As part of our ESO program, in the framework of the GRACE\footnote{{\sl GRB
Afterglow Collaboration at ESO}: see the web page \\ {\tt
http://www.gammaraybursts.org/grace/}} collaboration, we observed the
field of GRB000911 with both optical and NIR cameras. Our observations
started as soon as the precise position of the radio counterpart was made
public; we then followed up the optical/NIR afterglow for more than one
year with various telescopes located at ESO and at the Canary Islands
Observatories. In this paper we present our data and compare the results
with those published in Lazzati et al. (2001). In particular, since our
dataset is more complete in terms of time coverage and contains twice 
as many measurements as reported by those authors, we have a chance to 
test with higher confidence the presence of a SN. Moreover, our late-time 
observations allowed us to study in detail the broadband optical/NIR 
emission from the host galaxy of GRB000911.

Throughout this paper we will assume a cosmology with $H_0 =
65$~km~s$^{-1}$~Mpc$^{-1}$, $\Omega_{\Lambda} = 0.7$, $\Omega_{\rm m} =
0.3$. Throughout the text, unless otherwise stated, errors and upper
limits are at 1$\sigma$ and 3$\sigma$ confidence level, respectively.

\begin{figure*}
\parbox{12.0cm}{
\psfig{file=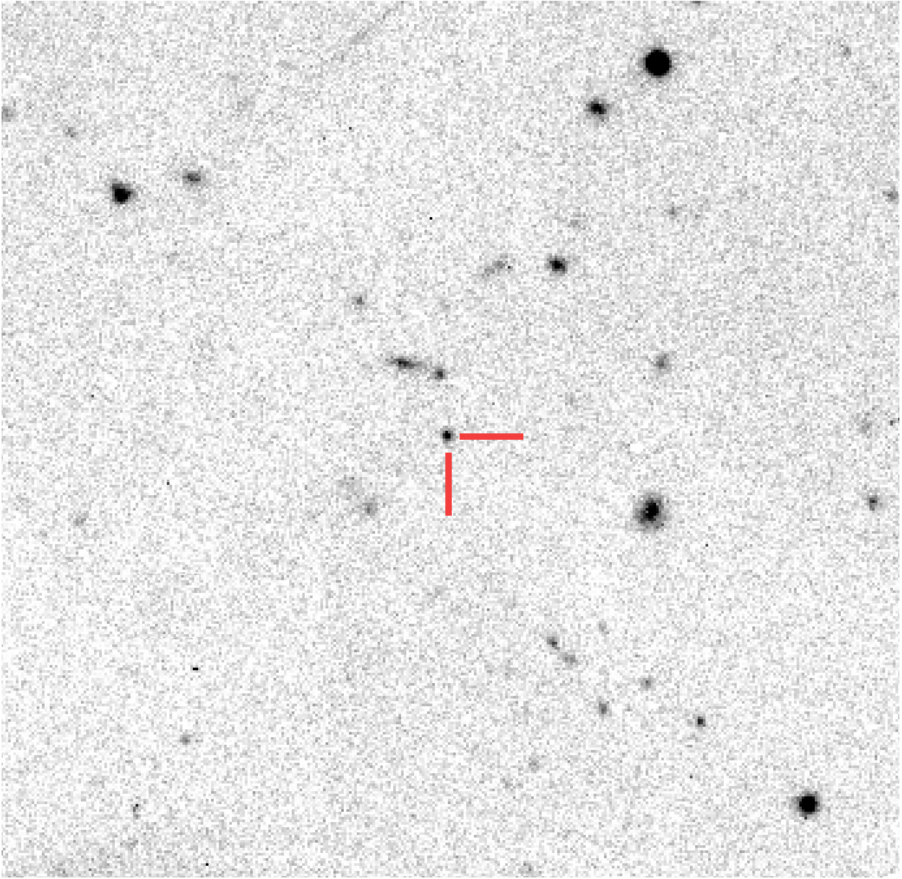,width=12cm}
}
\hspace{0.2cm}
\parbox{5.5cm}{
\vspace{8.0cm}
\caption{VLT-{\it Antu} $R$-band image of the afterglow of GRB000911
acquired on September 19, 2000. The field size is about 
1$\farcm$2$\times$1$\farcm$2; North is at top, East is to the left. 
The OT is indicated by the tick marks, and several faint extended objects 
can be seen around it. The limiting magnitude of the image is $R \sim$ 25.5.}
}
\end{figure*}

\begin{figure*}
\parbox{12.0cm}{
\psfig{file=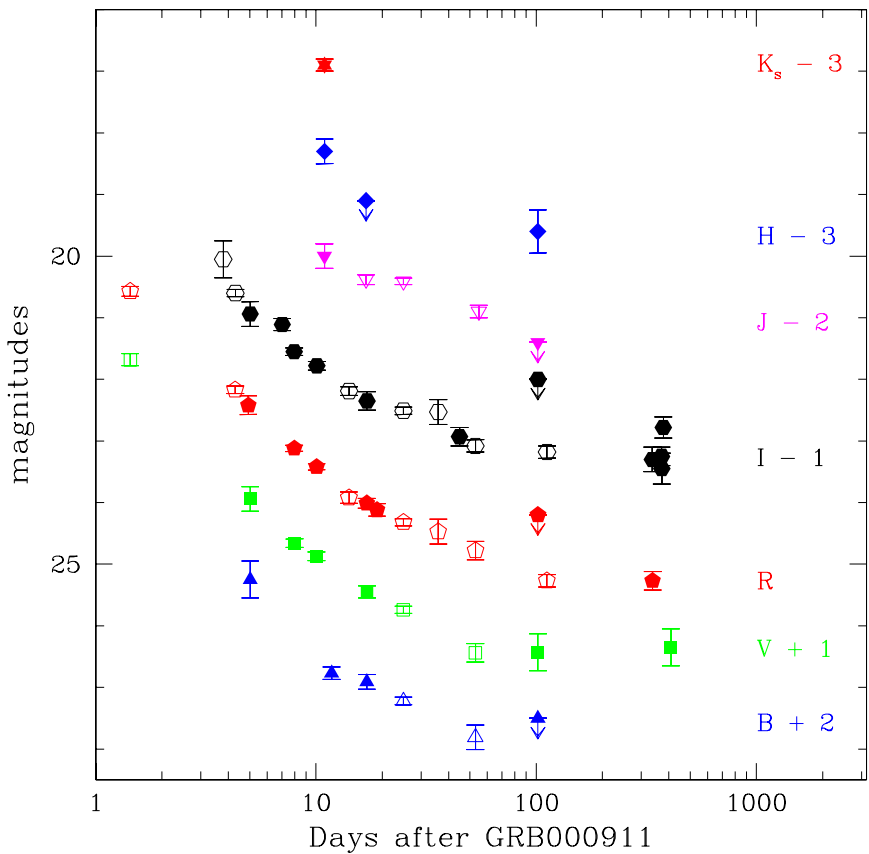,width=12cm}
}
\parbox{5.5cm}{
\vspace{7.5cm}
\caption{Light curves of the afterglow of GRB000911. Filled symbols
represent our data, open symbols refer to the data reported in Lazzati et
al. (2001) and Price et al. (2002). Different symbols correspond to 
different optical/NIR filters. The data are not corrected for Galactic 
extinction.}
}
\end{figure*}

\begin{table*}[t!]
\caption[]{Journal of the optical and NIR photometry of the GRB000911 
afterglow. Magnitude uncertainties are at 1$\sigma$ confidence level; upper 
limits are at 3$\sigma$ confidence level. The magnitudes are not corrected 
for Galactic interstellar extinction.}
\begin{center}
\begin{tabular}{rccccl}
\noalign{\smallskip}
\hline
\noalign{\smallskip}
\multicolumn{1}{c}{Mid-exposure} & Telescope & Filter & Total exposure &
Seeing & \multicolumn{1}{c}{Magnitude}\\
\multicolumn{1}{c}{time (UT)} & & & time (s) & (arcsec) &  \\
\noalign{\smallskip}
\hline
\noalign{\smallskip}

2000 Sep 16.328 & {\it Kueyen} & $B$ & 6$\times$1200 & 0.8 &
23.3$\pm$0.3$^*$\\ 
2000 Sep 23.075 & TNG          & $B$ &  3$\times$600 & 1.1 &
24.77$\pm$0.10\\ 
2000 Sep 28.326 & {\it Antu}   & $B$ &  5$\times$180 & 1.0 &
24.91$\pm$0.12\\
2000 Dec 22.163 & {\it Antu}   & $B$ &  4$\times$300 & 0.8 & 
$>$25.5\\
 & & & & & \\
2000 Sep 16.328 & {\it Kueyen} & $V$ & 6$\times$1200 & 0.8 &
22.9$\pm$0.2$^*$\\ 
2000 Sep 19.288 & {\it Antu}   & $V$ &  5$\times$180 & 0.7 &
23.66$\pm$0.07\\
2000 Sep 21.357 & {\it Antu}   & $V$ &  7$\times$180 & 0.7 &
23.87$\pm$0.07\\
2000 Sep 28.339 & {\it Antu}   & $V$ &  5$\times$180 & 1.0 &
24.45$\pm$0.1\\
2000 Dec 22.181 & {\it Antu}   & $V$ &  3$\times$300 & 0.9 &
25.43$\pm$0.3\\
2001 Oct 25.203 & {\it Antu}   & $V$ &  4$\times$300 & 0.6 &
25.35$\pm$0.3\\
 & & & & & \\
2000 Sep 16.235 & {\it Kueyen} & $R$ &  2$\times$60  & 0.8 &
22.42$\pm$0.15\\
2000 Sep 19.274 & {\it Antu}   & $R$ &  5$\times$180 & 0.6 &
23.12$\pm$0.05\\
2000 Sep 21.374 & {\it Antu}   & $R$ &  5$\times$180 & 0.8 &
23.42$\pm$0.05\\
2000 Sep 28.313 & {\it Antu}   & $R$ &  5$\times$180 & 0.8 &
24.01$\pm$0.08\\
2000 Sep 30.244 & {\it Antu}   & $R$ &  4$\times$60  & 0.8 &
24.12$\pm$0.10\\
2000 Dec 22.194 & {\it Antu}   & $R$ &  3$\times$300 & 1.0 & 
$>$24.2\\
2001 Aug 15.659 & NOT          & $R$ & 15$\times$900 & 1.1 &
25.27$\pm$0.15\\
 & & & & & \\
2000 Sep 15.097 & NOT          & $I$ &  4$\times$420 & 1.4 &
21.05$\pm$0.3\\
2000 Sep 16.328 & {\it Kueyen} & $I$ & 6$\times$1200 & 0.8 &
21.9$\pm$0.2$^*$\\ 
2000 Sep 18.333 & {\it Antu}   & $I$ &  1$\times$180 & 0.9 &
22.11$\pm$0.10\\
2000 Sep 19.261 & {\it Antu}   & $I$ &  5$\times$180 & 0.7 &
22.55$\pm$0.06\\
2000 Sep 21.388 & {\it Antu}   & $I$ &  5$\times$180 & 0.8 &
22.78$\pm$0.07\\
2000 Sep 28.353 & {\it Antu}   & $I$ &  5$\times$180 & 0.9 &
23.35$\pm$0.15\\
2000 Oct 26.177 & {\it Antu}   & $I$ &  5$\times$180 & 0.7 &
23.93$\pm$0.15\\
2000 Dec 22.208 & {\it Antu}   & $I$ &  3$\times$300 & 1.1 &
$>$23.0\\
2001 Aug 14.166 & NOT          & $I$ & 15$\times$600 & 0.9 &
24.3$\pm$0.2\\
2001 Sep 17.377 & {\it Antu}   & $I$ &  3$\times$360 & 0.9 &
24.25$\pm$0.15\\
2001 Sep 18.399 & {\it Antu}   & $I$ &  3$\times$360 & 0.8 &
24.45$\pm$0.25\\
2001 Sep 24.375 & {\it Antu}   & $I$ &  3$\times$360 & 0.6 &
23.78$\pm$0.17\\
 & & & & & \\
2000 Sep 22.230 & NTT          & $J$ &           900 & 0.9 &
22.0$\pm$0.2\\
2000 Dec 22.125 & {\it Antu}   & $J$ &          1800 & 0.6 &
$>$23.4\\
 & & & & & \\
2000 Sep 22.240 & NTT          & $H$ &           900 & 0.9 &
21.3$\pm$0.2\\
2000 Sep 28.200 & NTT          & $H$ &          3240 & 1.3 & 
$>$22.1\\
2000 Dec 22.063 & {\it Antu}   & $H$ &          5400 & 0.6 &
22.6$\pm$0.35\\
 & & & & & \\
2000 Sep 22.250 & NTT         & $K_s$ &          900 & 0.8 &
19.9$\pm$0.1\\
\noalign{\smallskip}
\hline
\noalign{\smallskip}
\multicolumn{6}{l}{$^*$This measurement was derived from a spectroscopic
observation}\\
\end{tabular}
\end{center}
\end{table*}

\section{Observations and data reduction}

We started our observational campaign on the OT of GRB000911 at ESO on
September 16, 2000. The complete log of our imaging observations is
reported in Table~1. 

\subsection{Optical photometry}

Optical $BVRI$ data were collected at Cerro Paranal (Chile) with VLT-{\it
Antu} plus FORS1 and with VLT-{\it Kueyen} plus FORS2 over a baseline of
more than 13 months. FORS1 and FORS2 were equipped with a Tektronix
and a SiTE CCD, respectively, in both cases with a 2048$\times$2048 pixel
array; both instruments covered a 6$\farcm$8$\times$6$\farcm$8 
field in the standard resolution imaging mode, giving a scale of 
0$\farcs$2 pix$^{-1}$.

A single $B$-band pointing was obtained on September 23, 2000, with
TNG+DOLoRes at La Palma, Canary Islands (Spain); the imaging spectrograph
DOLoRes carried a 2048$\times$2048 pixel Loral backside-illuminated 
CCD which images a field of 9$\farcm$5$\times$9$\farcm$5 with a scale of
0$\farcs$275 pix$^{-1}$. Early $I$-band and late-time $R$- and $I$-band
observations were acquired at NOT (La Palma, Canary Islands), with the
ALFOSC instrument. This also was equipped with a 2048$\times$2048 pixel
Loral CCD, giving a field of view of 6$\farcm$4$\times$6$\farcm$4 and an
image scale of 0$\farcs$188 pix$^{-1}$.

Optical images were bias-subtracted and flat-fielded with the standard
reduction procedure. In general, frames taken on the same night in the
same band were summed together in order to increase the signal-to-noise
ratio. 
We performed photometry with standard Point Spread Function (PSF) fitting 
using the {\sl DAOPHOT II}~image data analysis package 
PSF-fitting algorithm (Stetson 1987) within MIDAS\footnote{MIDAS 
(Munich Image Data Analysis System) is developed, distributed and 
maintained by ESO (European Southern Observatory) and is available at 
{\tt http://www.eso.org/projects/esomidas}}. The PSF-fitting 
photometry is accomplished by 
modeling a two-dimensional Gaussian profile with two free parameters (the 
half width at half maxima along $x$ and $y$ coordinates of each frame) on 
at least 5 unsaturated bright stars in each image. The errors associated 
with the measurements reported in Table 1 represent statistical 
uncertainties obtained with the standard PSF-fitting procedure.

For the late-epoch observations (i.e. those taken starting December 2000),
when the OT was faint and a substantial contribution from an underlying 
host galaxy became apparent (see Sect. 3.1),
we checked the results of the PSF-fitting algorithm by determining the
optical magnitudes of the object with the aperture photometry method. In
this case we used an aperture diameter equal to the Full Width at Half
Maximum (FWHM) of each summed image. Comparison between the results
obtained with the two methods indicates no appreciable difference within
the uncertainties. As we shall see below, this has implications for the
compactness and the morphology of the host of GRB000911.

To be consistent with optical magnitude measurements appearing on the GCN
circulars archive\footnote{GCN Circulars are available at:\\ {\tt
http://gcn.gsfc.nasa.gov/gcn/gcn3\_archive.html}}, the photometric
calibration was done
using the $BVRI$ magnitudes, as measured by Henden (2000), of several
field stars checked for their magnitude constancy. We find this magnitude
calibration to be accurate to within 3\%. However, this calibration is
offset with respect to the one used by Lazzati et al. (2001), depending on 
the considered band, by $\sim$0.1-0.2 mag. We have evaluated this
discrepancy by comparing our and their quasi-simultaneous ($<$2 hr)
measurements of September 28, 2000, and have corrected their photometry to
obtain a mutually consistent zero-point level. 
Because of the above procedure, we did not plot in Fig. 2 the 
measurements acquired by Lazzati et al. (2001) on that date.
It should also be noted that the photometry errors quoted throughout the
rest of the paper are only statistical and do not account for any possible
zero-point offset, which we expect to be smaller than 2\%.

\subsection{NIR photometry}

NIR imaging in $J$, $H$ and $K_s$ bands was obtained between September 21
and December 22, 2000, at the ESO NTT+SofI in La Silla (Chile), and again 
at the VLT-{\it Antu} in Paranal (Chile), equipped with ISAAC (see Table 1).
The SofI NIR camera carried a Rockwell Hawaii 1024$\times$1024 pixel HgCdTe 
array for imaging and spectroscopy in the 0.9--2.5 $\mu$m band. The plate 
scale was 0$\farcs$292 pix$^{-1}$ and the corresponding field of view was
4$\farcm$9$\times$4$\farcm$9. ISAAC was equipped, in the short-wavelength
(0.9--2.5 $\mu$m) NIR range, with a similar Rockwell Hawaii
1024$\times$1024 pixel HgCdTe array which had a scale of 0$\farcs$148
pix$^{-1}$ and a field of view of 2$\farcm$5$\times$2$\farcm$5. The $K_s$
filter is centered at 2.12 $\mu$m and has a FWHM of 0.34 $\mu$m. For each 
NIR observation the total integration time was split into images of 40 s 
each, with dithering after each individual exposure. 

Reduction of the NIR images was performed with the IRAF and STSDAS
packages\footnote{IRAF is the Image Analysis and Reduction Facility made
available to the astronomical community by the National Optical Astronomy
Observatories, which are operated by AURA, Inc., under contract with the
U.S. National Science Foundation. STSDAS is distributed by the Space
Telescope Science Institute, which is operated by the Association of
Universities for Research in Astronomy (AURA), Inc., under NASA contract
NAS 5--26555. It is available at {\tt http://iraf.noao.edu/}}. 
Each image was reduced by first subtracting a mean sky, obtained from the 
median of a number of frames acquired just before and after each processed 
image. Before frames were used for sky subtraction, stars
in them were eliminated by a background interpolation algorithm ({\tt
imedit}) combined with an automatic ``star finder'' ({\tt daofind}). 
Then, a differential dome flatfield correction was applied to the 
sky-subtracted image, and the frames were registered to fractional pixels 
and combined. The telescope dithering was measured from the offsets of field 
objects in each image and the images were averaged together using 
inter-pixel shifts. Magnitudes were measured inside circular apertures 
of diameter comparable to the FWHM of each image.

We calibrated the NIR photometry with stars selected from the NICMOS
Standards List (Persson et al. 1998). The stars were observed in five
positions on the detector, and were reduced in the same way as the source
observations. Formal photometric accuracy based only on the standard star
observations is typically better than 3\%. We checked the consistency of
our $J$-band calibration with that determined by Lazzati et al. (2001):
we found that the two zero points agree well within the uncertainties.

\subsection{Optical Spectroscopy}

Two series of 6 spectra each, with individual exposure times of 20 
minutes, were acquired on September 16.328 and 30.301, 2000 (UT times at
mid-exposure), with VLT-{\it Kueyen}+FORS2 and VLT-{\it Antu}+FORS1,
respectively. The spectra were acquired using FORS Grism \#150I, which
nominally covers the 3300-9000 \AA~range, and with a slit width of 1$''$.
The use of a separation filter was needed to avoid overlapping of spectral
orders over a given wavelength; this reduced the spectral range of the two
spectra to 3800-8600 \AA. This setup secured a final dispersion of 5.5
\AA~pix$^{-1}$ for the two spectra.

The spectra, after correction for flat-field and bias, were background
subtracted and optimally extracted (Horne 1986) using IRAF.
Helium-Neon-Argon lamps were used for wavelength calibration; both spectra
were then flux-calibrated by using the spectroscopic standard EG21 (Hamuy
et al. 1992, 1994). Finally, spectra taken within the same observing run
were averaged together to increase the S/N ratio. The accuracy of the
wavelength and flux calibrations was checked against the position of night
sky lines and the photometric data collected around the epoch in which the
spectra were acquired, respectively. The typical error was 0.5 \AA~for the
wavelength calibration and 10\% for the flux calibration. Unfortunately,
the proximity of the Moon to the OT in the first spectroscopic run and the
faintness of the OT in the second one severely limited the quality of
these spectra. Thus, no significant spectral features were found, but we
were able to extract rough estimates of broadband $BVI$ magnitudes (see 
Table~1).

\section{Results}

\subsection{GRB000911 optical-NIR afterglow light curves}

In Fig. 1 a VLT image of the GRB000911 field is shown. The OT variability
and the consistency of its position with both the IPN error box and the
radio transient support the afterglow nature of the source indicated by the
tick marks. In this deep image (limiting magnitude: $R \sim$ 25.5), several 
faint extended objects can be seen around the OT.
In Fig. 2 we plot both our photometric measurements and those
presented in Lazzati et al. (2001), plus two late $R$ and $I$
points\footnote{These two points, having been calibrated following 
Lazzati et al. (2001), were also corrected for the zero-point offset 
described in 
Sect. 2.1.} obtained in January 2001 by Price et al. (2002). In the optical
bands, the initial decline slows down at about 20 days after GRB,
probably due to the underlying host galaxy, or possibly to the presence of a
supernova, as proposed by Lazzati et al. (2001). The NIR light curves are
not well sampled, but the $J$ and $H$ band observations also show a fading.
We also note (see Fig. 2 and Table 1) that at late times, $\sim$12 months
after the GRB, the $I$-band light curve shows a marginal increase of
$\sim$0.6 mag in 1 week. Due to this behaviour, we excluded the last $I$
point from our light curve fits described below. Possible interpretations
of this behaviour are briefly discussed in Sect. 4.1.
The $R$ and $I$ observations acquired with VLT-{\it Antu} on 
December 22, 2000 ($\sim$100 days after the GRB trigger) were performed 
under relatively poor seeing conditions and only provide loose upper 
limits to the OT magnitudes.

We evaluated the Galactic extinction in the optical and NIR
bands in the direction of GRB000911 using the Galactic dust infrared
maps by Schlegel et al. (1998); from these data we obtained a color
excess $E(B-V)$ = 0.120. By applying the relation of Cardelli et al.
(1989), we derived $A_B$ = 0.51, $A_V$ = 0.38, $A_R$ = 0.32, $A_I$ =
0.23, $A_J$ = 0.13, $A_H$ = 0.07 and $A_{K_s}$ = 0.04. The corrected
magnitudes were
then converted into flux densities following Fukugita et al. (1995)
for the optical filters and Bersanelli et al. (1991) for the NIR ones. We
also added a 5\% error in quadrature to the uncertainties on the
optical-NIR flux densities in order to account for differences in the
instrumental responses of different telescopes in the same band.

The optical and NIR colors of the OT of GRB000911 during the first 10 days
after the high-energy event fall in the loci populated by GRB afterglows
in the color-color diagrams as illustrated by \v{S}imon et al. (2001) and
by Gorosabel et al. (2002).

\subsection{The GRB000911 host galaxy}

\begin{figure}
\psfig{file=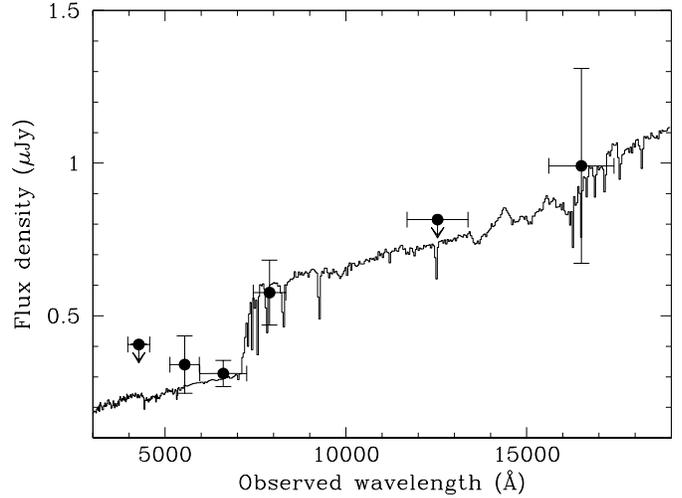,width=9cm,angle=-90}
\caption{$BVRIJH$ SFD of the host galaxy of GRB000911 corrected for
Galactic extinction (filled dots), superimposed onto one of the best-fit
templates of Table 2 (a starburst-like galaxy plus intrinsic reddening) 
obtained applying the {\sl Hyperz} code to the SFD data of the GRB000911 
host. The horizontal bars associated with the SFD data points indicate the 
FWHM of the photometric filters.}
\end{figure}

The host galaxy of GRB000911 appears unresolved in all of our VLT frames,
as the PSF of the galaxy image is always consistent with that of a
pointlike object. This implies an upper limit of 3.3 kpc on the host
galaxy half-light radius. To estimate the flux of the GRB000911 host, we
considered the measurements or upper limits acquired since December 2000
as reported in Table 1. In case of detections, when more than one
measurement was available in the same band, we considered their 
average value.

The fit to the GRB000911 host galaxy Spectral Flux Distribution
(SFD) was carried out with the use of the version 1.1 of the 
{\sl Hyperz}\footnote{available at: {\tt
http://webast.ast.obs-mip.fr/hyperz/}} code (Bolzonella et al. 2000a).
This allows a $\chi^2$ fit of the observed data with 8 different templates
of synthetic galaxy spectra (starburst, elliptical, lenticular, four kinds
of spirals, and irregular). In all cases, the evolution of the Star
Formation Rate (SFR) in the galaxy templates is modeled using an
exponential law. The Initial Mass Function is modeled assuming the law 
by Miller \& Scalo (1979). In the fits a solar metallicity $Z = Z_\odot 
\simeq 0.02$ (with $Z$ the mass fraction of heavy elements) was assumed.

This code also allows an estimate of the photometric redshift of the
studied galaxy, the age of its dominant stellar population, and the
presence of further overall absorption local to the galaxy itself. Four
possible extinction laws were considered, i.e., those by Seaton (1979),
Fitzpatrick (1986), Pr\'evot et al. (1984) and Calzetti et al. (2000)
which are suitable to describe the extinctions within the Milky Way, the
Large Magellanic Cloud (LMC), the Small Magellanic Cloud (SMC) and a
generic starburst (Stb) galaxy, respectively.

By considering the templates over a redshift range $z$ = $0-5$ with a step
$\Delta z$ = 0.001, and varying the $V$-band local extinction $A_V$ in 
the range $0-2$ with $\Delta A_V$ = 0.01, we find that the template which
best fits our VLT $VRIH$ data points and $BJ$ upper limits of the
GRB000911 host is that of an irregular galaxy at $z$ = 0.904 with the
presence of slight reddening due to either a LMC-, or a SMC-, or a
Stb-like absorption law (see Table 2 and Fig. 3).
The spectroscopically determined redshift ($z$ = 1.0585$\pm$0.0001;
Price et al. 2002) is within the 68\% confidence level interval of that 
obtained from our photometry using the {\sl Hyperz} code (see Table 2). 
The best-fit age of the host galaxy stellar population is 0.7 Gyr for a 
SMC-like template and 0.5 Gyr for LMC- and Stb-like templates.
Extinction is moderate with $A_V\,=\,0.3-0.5$.

\begin{table*}
\caption[]{Best-fit parameters of the GRB000911 host galaxy obtained
with the {\sl Hyperz} code. Photometric redshift errors are at 68\%
confidence level. The value of the luminosity $L^\star$ at the knee of the 
galaxies luminosity distribution (as defined by Schechter 1976; see 
Sect. 4.3) is taken from Lilly et al. (1995).}
\begin{center}
\begin{tabular}{ccccccc}
\hline
\hline
Template & $\chi^2$/dof & Photometric & Age   & $A_V$ & M$_B$ & $L/L^\star$ \\
galaxy   &              & redshift    & (Gyr) & (mag) &       &         \\
\noalign{\smallskip}
\hline
\noalign{\smallskip}

Irr-LMC & 0.685 & 0.904$^{+0.125}_{-0.128}$ & 0.509 & 0.39 & $-$18.40 & 0.066 \\
Irr-Stb & 0.731 & 0.904$^{+0.134}_{-0.130}$ & 0.509 & 0.55 & $-$18.38 & 0.066 \\
Irr-SMC & 0.732 & 0.904$^{+0.149}_{-0.138}$ & 0.719 & 0.32 & $-$18.37 & 0.064 \\

\noalign{\smallskip}
\hline
\noalign{\smallskip}
\end{tabular}
\end{center}
\end{table*}

The assumption of an evolving metallicity, as remarked by several
authors (e.g., Bolzonella et al. 2000a,b; Gorosabel et al. 2003a,b;
Christensen et al. 2004), can be considered as a second-order parameter 
of the fit; thus, this is not expected to significantly alter our results 
and conclusions. The same applies to the choice of a LMC- or a SMC-like 
template, instead of a starburst, for the host galaxy.

Using the best-fit galaxy template we have computed the expected host flux
in the $B$, $J$ and $K_s$ band, for which we only have either upper limits
or no late-time observations. The magnitudes, not corrected for Galactic
extinction, are $B$ = 26.1, $J$ = 23.6 and $K_s$ = 21.9. For each of these 
three values an uncertainty of 0.3 mag can be assumed.

\begin{figure*}[th!]
\parbox{10.0cm}{
\vspace{-3.5cm}
\psfig{file=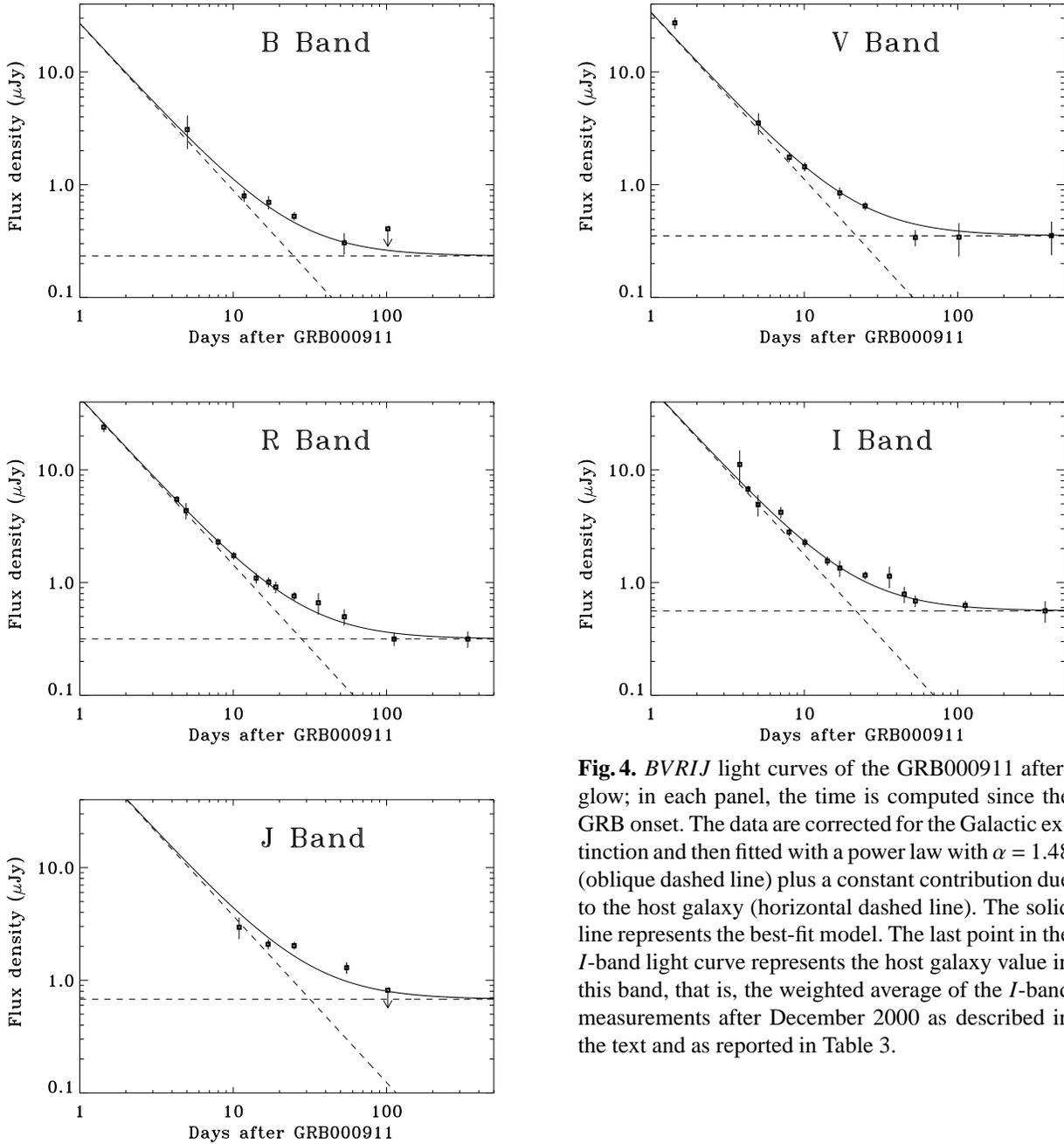,width=18cm}
}
\hspace{-8.8cm}
\parbox{7.5cm}{
\vspace{10.5cm}
\caption{$BVRIJ$ light curves of the GRB000911 afterglow; in each panel, the 
time is computed since the GRB onset. The data are corrected for the 
Galactic extinction and then fitted with a power law with $\alpha$ = 1.48 
(oblique dashed line) plus a constant contribution due to the host galaxy 
(horizontal dashed line). The solid line represents the best-fit 
model. The last point in the $I$-band light curve represents the host galaxy 
value in this band, that is, the weighted average of the $I$-band 
measurements after December 2000 as described in the text and as reported 
in Table 3.} 
}
\end{figure*}

\begin{figure*}[th!]
\parbox{10.0cm}{
\vspace{-6.5cm}
\psfig{file=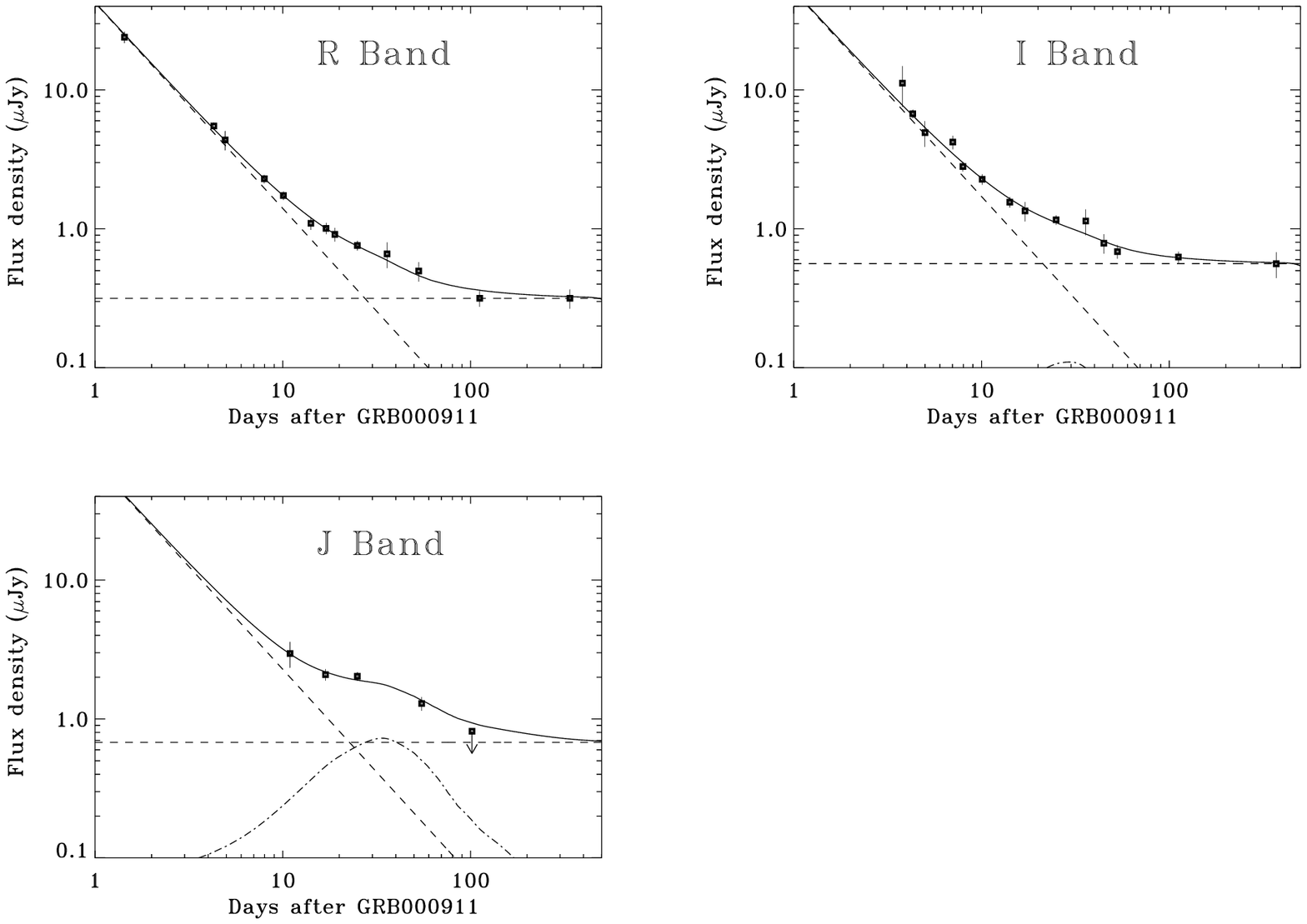,width=18cm}
}
\hspace{-8.8cm}
\parbox{7.5cm}{
\vspace{4cm}
\caption{Same as Fig. 4 for the $RIJ$ light curves of the GRB000911 
afterglow, but with the addition of a SN1998bw at $z$ = 1.06 (dot-dashed 
line; in the $R$-band panel this is not visible) brightened, with respect to 
the original one, by the factors reported in Table 3. The solid line 
represents the best-fit model. In these fits the SN explosion time was 
assumed to coincide with the GRB start time.}
}
\end{figure*}

\subsection{GRB000911 afterglow light curve fits}

In order to study the GRB000911 optical/NIR afterglow light curves and to 
determine the presence of an underlying SN contribution, we took into 
account the host galaxy contamination as measured at late epochs with VLT 
($VRIH$ bands) or as inferred through the {\sl Hyperz} fits ($BJK_s$ bands).
The host magnitudes and upper limits were also corrected for Galactic 
reddening and converted to flux densities in the same way applied to the 
afterglow magnitudes as described in Sect. 3.1.

We then fitted each of the $BVRIJ$ light curves with a power-law plus
a constant (the latter simulating the galaxy contribution). The results,
reported in the upper part of Table 3, show that throughout the optical
range, the decay slope is consistent with a constant value. As one
can see from the $\chi^2$ values, the optical light curves are well
described by this model with no need for any further component. The
$J$-band data instead appear to have a substantially flatter decay
($\alpha_J$ = 0.79$\pm$0.15) and an unacceptable fit ($\chi^2$/dof = 2.6,
where `dof' stands for `degrees of freedom').

However, given that the $J$-band light curve is not sampled in the first
10 days after the GRB, we adopted the same decay slope for the $BVRIJ$
bands assuming the weighted average decay index found in the optical
bands: this resulted in $\alpha_{\rm opt}$ = 1.48$\pm$0.02. The fit
results, illustrated in the central part of Table 3 and in Fig. 4, are
again basically acceptable for the optical light curves, except for the
$J$-band. Therefore, following Lazzati et al. (2001), we added the
prototypical `hypernova' SN1998bw (Galama et al. 1998; Patat et al. 2000)
redshifted to $z$ = 1.0585 (Djorgovski et al. 2001; Price et al. 2002) and
dimmed accordingly; while fitting the light curve in each band, we also
left the SN flux normalization vary as a free parameter. Details on this
procedure can be found, e.g., in Appendix A of Zeh et al. (2004).

Our approach is more general than that of Lazzati et al. (2001); while
those authors consider a collimated post-break fireball evolution
which requires that the decay index be twice the spectral slope (as
in Sari et al. 1999), we do not make any a priori assumption about the
fireball hydrodynamics. We also have not tried to estimate a possible
supernova contribution to the $B$ and $V$ observed light curves, because
there are no ultraviolet (UV) data for SN1998bw to be used as templates to
be transformed to those observed bands. Finally, we chose not to
simulate UV light curves via extrapolation of the optical spectrum owing
to the unpredictably faster variability of the UV flux with respect to the
optical flux in supernovae (Cappellaro et al. 1995).

The lower part of Table 3 shows that this model produces a good
description of the $RIJ$ dataset (see Fig. 5). 
The fit improvement obtained by including the presence of a SN, as
measured by applying the $F$-test procedure (e.g., Press et al. 1992), is
real at a confidence level of 92\%. The difference between the results by
Lazzati et al. (2001) and the present ones can be explained partly by our
much more accurate estimate of the host galaxy $J$-band magnitude, and
partly by the fact that those authors did not fully take into account the
limitations in the applicability of the $F$-test (see Protassov et al.
2002), namely, the interdependence among fit parameters. Indeed, 
considering these limitations, one obtains an actual significance of 
$\sim$85\% for the presence of a SN in the GRB000911 light curves
presented by Lazzati et al. (2001).

Our fits are thus consistent with a SN at a flux comparable to that of SN 
1998bw, having exploded simultaneously (with an uncertainty of about one 
week) with GRB000911, thus confirming the findings of Lazzati et al. 
(2001). Possibly, a slight preference (though not statistically 
significant) for the SN having exploded $\sim$1 week before the GRB is 
implied by the $J$-band fits.
We remark that the fits reported in the lower part of Table 3 and illustrated 
in Fig. 5 are made assuming simultaneity between the GRB and SN explosions.

We also attempted fits with different Type Ia and Ic SN templates, chosen 
among those which have well-sampled light curves from the earliest epochs.
Specifically, we used SNe 2002ap (Gal-Yam et al. 2002; Pandey et al. 2003),
1994D and 1994I (Barbon et al. 1999 and references therein) and obtain,
in terms of statistical confidence, results similar to 
those derived with SN1998bw.

The $H$-band light curve has not been fitted, because of insufficient
sampling. The $H$-band upper limit would suggest a much steeper decay
than observed in the other bands. Although we have no reason to believe
that the upper limit is not reliable, owing to the scarcity of
measurements in this band we do not speculate about this faster decay.

\begin{table*}
\caption[]{Best-fit parameters of the GRB000911 afterglow light curve 
in the $BVRIJ$ bands, computed assuming a power law decay plus
a constant (host galaxy) contribution with and without the presence
of an underlying 1998bw-like SN occurring simultaneously with the GRB. 
Host galaxy magnitudes reported here are not corrected for Galactic 
reddening. Values within brackets are fixed in the corresponding fit.}
\begin{center}
\begin{tabular}{ccccc}
\hline
\hline
Filter & $\alpha$ & Host galaxy magnitude & SN intensity$^a$ & 
$\chi^2$/dof \\
\hline
\multicolumn{5}{c}{Fits without SN plus free $\alpha$}\\
\hline
$B$ & $1.12 \pm 0.17$ & [26.1]  & --- & 4.2/4 \\
$V$ & $1.63 \pm 0.05$ & [25.4]  & --- & 7.3/7 \\
$R$ & $1.45 \pm 0.03$ & [25.27] & --- & 7.4/11 \\
$I$ & $1.43 \pm 0.06$ & [24.3]  & --- & 12.6/12 \\
$J$ & $0.79 \pm 0.15$ & [23.6]  & --- & 7.9/3 \\
\hline
\multicolumn{5}{c}{Fits without SN plus fixed $\alpha$}\\
\hline
$B$ & [1.48] & [26.1]  & --- & 8.3/5 \\
$V$ & [1.48] & [25.4]  & --- & 14.0/8 \\
$R$ & [1.48] & [25.27] & --- & 8.4/12 \\
$I$ & [1.48] & [24.3]  & --- & 13.2/13 \\
$J$ & [1.48] & [23.6]  & --- & 28.4/4 \\
\hline
\multicolumn{5}{c}{Fits with SN1998bw plus fixed $\alpha$}\\
\hline
$R$ & [1.48] & [25.27] & 3.5$\pm$2.0 & 5.8/11 \\
$I$ & [1.48] & [24.3]  & 3.4$\pm$1.7 & 8.8/12 \\
$J$ & [1.48] & [23.6]  & 0.9$\pm$0.3 & 3.6/3 \\
\hline
\multicolumn{5}{l}{$^a$ Ratio between the SN 1998bw light peak and
that of the best-fit SN.}\\

\end{tabular}
\end{center}
\end{table*}

\subsection{GRB000911 optical-NIR SFDs}

To study the broadband spectral variability of the afterglow, we
subtracted the host galaxy spectrum from the observed fluxes in optical
and NIR bands at five epochs after the GRB, at which the sampling has
the widest spectral coverage. In the cases in which the data in
different bands were not simultaneous, the flux density was
interpolated to the reference epoch by using the best-fit power-law
decay $\alpha$ = 1.48. Magnitudes were corrected for Galactic reddening
and converted into flux densities following the procedure already
described.

The simultaneous afterglow SFDs thus obtained, not further corrected for
the intrinsic extinction within the host galaxy, are reported in Fig. 6.
Despite the uncertainties (due to the rather large errors on the host
galaxy flux densities), a spectral steepening with time is suggested.
Assuming a spectrum $F(\nu) \propto \nu^{-\beta}$ for all epochs, the
$V-R$ color at 1.4 days after the GRB implies a spectral slope $\beta$ =
$-$0.7$^{+3.4}_{-3.2}$. A fit to the broad-band optical/NIR SFDs at the
following 4 epochs yields $\beta$ = 0.8$\pm$1.3 (5.03 days after
GRB000911), $\beta$ = 1.1$\pm$0.3 (10.94 days), $\beta$ = 1.0$\pm$1.2
(16.90 days) and $\beta$ = 1.9$\pm$0.7 (25.00 days). The results obtained
for the epochs between $\sim$5 and $\sim$17 days after the GRB trigger are
broadly consistent with those derived by Lazzati et al. (2001).
We remark that the huge errors on the value of $\beta$ referring to the 
first epoch (September 12.74) are due to two effects: the moderately large 
uncertainties ($\sim$0.1 mag) on the magnitudes and the very narrow 
($\sim$0.1 dex) SFD baseline.

\begin{figure}
\psfig{file=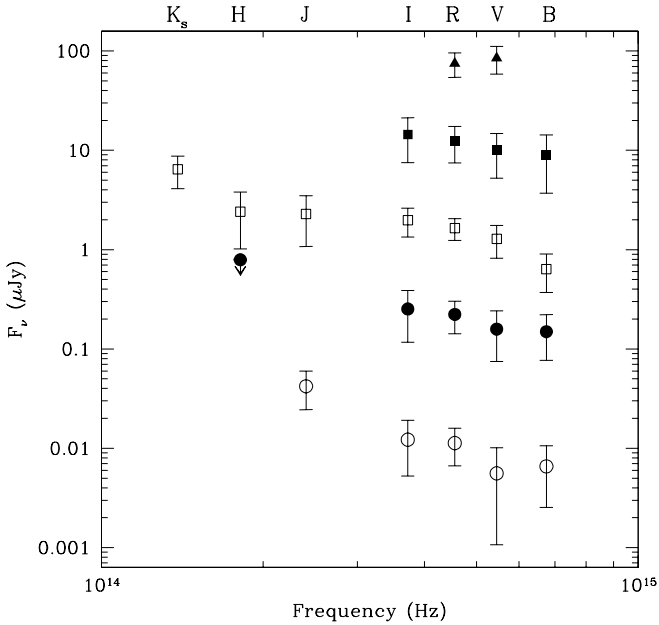,width=9cm}
\vspace{-.5cm}
\caption{SFDs of the afterglow of GRB000911 at 1.44 (September 12.74, 
filled triangles), 5.03 (September 16.33, filled squares), 10.94 
(September 22.24, empty squares), 16.90 (September 28.20, filled circles) 
and 25.00 (October 6.30, open triangles) days after GRB. The SFDs of 
September 12.74 and 16.33 are shifted up by 0.5 dex, while those of 
September 28.20 and October 6.30 are shifted down by 0.5 and 1.5 dex, 
respectively. The data are corrected for Galactic extinction and for the 
flux of the host galaxy.}
\end{figure}

If we correct the afterglow data for the absorption as derived from the
{\sl Hyperz} best fits of the host galaxy SFD, we obtain for the OT the
spectral slopes $\beta$ = $-$1.3$\pm$3.3, 0.2$\pm$1.3, 0.6$\pm$0.3,
0.4$\pm$1.2 and 1.3$\pm$0.7 at the 5 considered epochs. No substantial 
difference is found among the considered reddening laws (Stb, SMC and LMC; 
see also Kann 2004).

\section{Discussion}

Our late-time observational campaign on GRB000911 has allowed us, for the
first time, to directly observe and thus extract important information on
its host galaxy; this in turn has helped us to better model the shape and
evolution of the light curves and SFDs of the GRB000911 afterglow in the
optical/NIR domain. In the following we discuss the results obtained on
each of these three issues.

\subsection{The GRB afterglow}

The temporal decay index of the afterglow of this GRB is consistent with 
the decline being wavelength-independent. In the first week, the spectral 
slope $\beta \sim 1$ (assuming no intrinsic absorption local to the host),
combined with the average temporal index $\alpha \sim 1.5$, is consistent
with a spherical fireball expansion into a constant-density circumburst
medium under the hypothesis that the cooling break frequency of the
synchrotron radiation, $\nu_{\rm c}$, is above the optical range (Sari et
al. 1998). This implies an electron energy distribution index $p\sim$ 3,
to be compared with the range 2--2.5 found for the $p$ values of GRBs
studied by Frontera et al. (2000). It should however be remarked that
Masetti et al. (2000)  suggested that $p >$ 2.6 for the afterglow of
GRB990705. Unfortunately, the lack of information concerning the
high-energy afterglow of GRB000911 does not allow us to explore in detail
more complex models for the fireball expansion. Price et al. (2002),
assuming that a light curve break due to collimated emission (as described
in Sari et al. 1999) occurred $\sim$1 day after the GRB000911 onset,
derived $p$ = 1.5.

The spectral index at 25 days after the GRB, $\beta \sim 1.9$, is again
not supportive of a jet geometry of the afterglow emission in the
classical collimated fireball scenario (Sari et al. 1999), unless the
light curve changes slope at late epochs, possibly implying a more
significant supernova contribution. The data are not sufficient to test a
change of temporal slope of the afterglow.

A collimated fireball before the jet break (Sari et al. 1999;
Rhoads 1999) and an isotropic fireball (Sari et al. 1998) are described in
equivalent ways by the standard model, and they cannot be discerned by
the temporal and spectral indices. As noted for the case of GRB020405
(Masetti et al. 2003), if the bump observed in the optical-NIR light
curves is due to an emission component independent of the afterglow
(i.e., a SN), our data cannot rule out a steepening of the afterglow light
curve, related to a collimation break, occurring more than $\sim$10 days
after the GRB. According to Sari et al. (1999), this would imply a lower
limit to the jet opening angle of $\theta_0 > 10^\circ$ assuming a
circumburst density $n$ = 1 cm$^{-3}$ and a GRB isotropic equivalent
energy $E_{\rm iso}$ = 7.8 $\times$10$^{53}$ erg (Price et al. 2002).

A ``flat $p$" distribution (i.e. with $p <$ 2), as suggested by Price et 
al. (2002) for this afterglow, may instead be invoked if local host 
absorption, as per the {\sl Hyperz} fit results, is considered.
However, in this case the fireball expansion (in the model of Dai 
\& Cheng 2001) is only compatible with a collimated model, assuming 
a very flat ($p <$ 1) electron distribution and $\nu_{\rm c}$ below 
the optical-NIR bands, even accounting for the large errors on $\beta$.

The ``inverted" spectral index hinted by the $VR$ data at day 1.44 
after the GRB, because of its large uncertainties due to its very short 
spectral baseline, is in agreement with the fireball model description.

Concerning the very last point in the $I$-band light curve (see
Fig. 2 and Table 1), we have no clear explanation for this behaviour, as
neither spectral nor color information is available for it, and the
temporal sampling is extremely poor. Having excluded any spurious effect
as the cause of this variability, for a transient at this redshift we are
left with two possibilities: a SN or an Active Galactic Nucleus. The host
galaxy optical spectrum (Price et al. 2002), however, does not support the
latter explanation. Thus, it could be that a SN, completely
unrelated to GRB000911, exploded in this same galaxy $\sim$6 rest-frame
months after the GRB. Indeed, given that the host is unresolved in the VLT
frames, its star forming regions, and thus any possible transient
associated with them, fall within the PSF of the galaxy image. A variable
object (possibly a SN) close in the sky to the GRB but unrelated to it 
was discovered in the field of GRB020405 (Masetti et al. 2003), and
two SNe (2002ap and 2003gd) have recently exploded in the nearby galaxy
M74 (Hirose, reported by Nakano 2002; Evans 2003) separated in time by
$\sim$1 year. In any case, given the low significance of the detection and
the scanty temporal sampling of the phenomenon, we regard this explanation
as tentative at best.

\subsection{The associated SN}

The results of the fits to the optical and NIR light curves (Table 3)
suggest that the presence of a SN underlying the afterglow of GRB000911 is
not required by the optical dataset, which is the best sampled, while it
is needed at 92\% confidence level to explain the $J$-band data under the
assumption that the decay in the NIR range is the same as in the optical.
Therefore, it seems that a SN similar to SN1998bw does not contribute
significantly to the (observer's frame) optical afterglow of GRB000911.

As one can see, the SN contribution in the NIR appears to be larger than,
although marginally consistent with, that in the optical bands.
This marginal discrepancy between the optical and the $J$ light curves 
might possibly be due in part to the presence of intrinsic reddening local 
to the host, as pointed out by the {\sl Hyperz} fit results. Indeed, the 
observer's frame optical bands correspond to UV and $B$ ranges at 
$z$ = 1.06 and are more strongly affected by absorption than the
observer's frame $J$ band.
A similar discrepancy between rest-frame UV and optical behaviours was 
found in the case of the possible SN associated with GRB020405 (Masetti 
et al. 2003), where the observed $V$-band data (which corresponded to 
the UV range at the redshift of this GRB, $z$ = 0.69) remained 
systematically below the best-fit SN curve.

We must however note, as already remarked by Price et al. (2003b), that
SNe Ib/c have an intrisic dispersion in the peak luminosity (see
Clocchiatti et al. 2000); therefore, in the present case, a SN fainter
than 1998bw could have been associated with GRB000911. A similar conclusion
was reached by Zeh et al. (2004) by considering the $R$-band light curves
of 21 GRB afterglows detected up to the end of 2002 and by studying the
presence of a possible SN contribution to the total light. Similarly, the
SN behaviour in the blue part of the spectrum, especially in the UV range,
is still poorly studied and one can not exclude that it may show
substantial differences from case to case.

Alternative explanations may also be considered, such as the model developed
by Beloborodov (2003) in which the fireball is interacting with a trailing
neutron shell; or that in which a SN remnant, located around the GRB
progenitor and excited by the GRB itself, cools down on time scales of weeks
(Katz 1994; Dermer 2002); or the dust echo model (Esin \& Blandford 2000;
Reichart 2001). Ramirez-Ruiz et al. (2001) modeled the GRB000911 data set of
Lazzati et al. (2001) considering the effect of fireball propagation through
the progenitor wind. However, the evidence indicating that (at least some)
long GRBs are associated with SNe (Galama et al. 1998; Della Valle et al.
2003; Stanek et al. 2003; Hjorth et al. 2003; Malesani et al.  2004) suggests
that these alternative models are less viable than the SN interpretation.

In conclusion, our late-epoch observations have allowed us to determine 
the host galaxy SFD and therefore to give a more accurate description of 
the GRB000911 OT light curves than was previously possible. 
With these data we could also pinpoint the
cause of the OT color evolution stressed by Lazzati et al. (2001): the OT
reddening at late epochs is mainly due to the presence of an underlying
host galaxy, redder than the afterglow, in the $R$ and $I$ filters, and by
a flux enhancement due to an additional (possibly a SN) component in the $J$
band. If this phenomenon is indeed produced by the emergence of a
Ib/c-type SN component associated with GRB000911, this is the farthest SN
connected with a GRB and, to the best of our knowledge, the farthest
core-collapse SN observed up to now.

\subsection{The host}

The temporal extension of our monitoring has allowed us to set some
constraints on the host galaxy of GRB000911. Its spectrum appears to be
compatible with that of an irregular galaxy with a moderate amount of
intrinsic extinction following a Magellanic-type or a Stb-like reddening
law, in agreement with the findings of Lazzati et al. (2001). The host
size is very compact: given that the galaxy is unresolved in all of our
late-epoch images, we estimated an upper limit of 3.3 kpc for its
half-light radius. This prevents us from studying the host morphology 
with our ground-based observations.

For the host we obtain an absolute $B$-band magnitude M$_B \sim$
$-$18.4, which compares acceptably well with the values $-$18.8 to
$-$18.9 obtained for this object by Price et al. (2002) and Le Floc'h et
al. (2003). This value is typical of host galaxies of GRBs (see, e.g., Le
Floc'h et al. 2003, Christensen et al. 2004 and Sokolov et al. 2001); in
particular, the GRB000911 host falls roughly in the middle of the 
absolute $B$-band magnitude distribution of the host galaxy sample of Le 
Floc'h et al. (2003; see their Fig. 6).

In terms of a comparison between the luminosity of the GRB000911 host
galaxy with that of the knee of the galaxies luminosity function,
$L^\star$ (Schechter 1976), we find (see Table 2) that $L_{\rm host}$
$\sim$ 0.07$L^\star$ using the luminosity function of Lilly et al. (1995).
If we use the luminosity function of Schechter (1976) we get values of
$L_{\rm host}$ which are about twice, in units of $L^\star$, those
obtained from that of Lilly et al. (1995), as noticed by Gorosabel et al.  
(2003a). Thus, averaging the two results, we can confidently say that the
host of GRB000911 has $L \sim$ 0.1$L^\star$ and is therefore a subluminous
galaxy. This is consistent with the findings of Hogg \& Fruchter (1999)
who found that GRB hosts are in general subluminous galaxies.

Using the best-fit templates decribed above and the relation between the
rest-frame luminosity at 2800 \AA~as reported in Kennicutt (1998) we can
give an estimate of the SFR in the GRB000911 host galaxy: from the fits
we get a SFR (corrected for the local host absorption) of $\sim$2.7
$M_\odot$ yr$^{-1}$ for all cases reported in Table 2. This again is
consistent with the SFR value obtained by Price et al. (2002) from the
[O{\sc ii}] $\lambda$3727 line of the host; however, it is slightly
different from the value those authors obtained using the 2800 \AA~flux
method. The difference can be attributed to the local absorption we found
in our analysis (see also Price et al. 2002 and Lazzati et al. 2001).
This further local absorption has been invoked to explain `dark' GRBs
(such as GRB000210; Piro et al. 2002) occurring in host galaxies with low
global extinction (Gorosabel et al. 2003a).

Our SFR estimate for the GRB000911 host, together with a luminosity $L
\sim$ 0.1$L^\star$, implies a specific SFR of $\approx$30 $M_\odot$ yr$^{-1}$
($L/L^\star$)$^{-1}$. This is a very high value: it exceeds by a factor
$\sim$3 the highest specific SFR in the GRB host sample of Christensen et
al. (2004), and is an outlier at the high end of the specific SFR
distribution of the Hubble Deep Field North (whose galaxies have specific
SFRs $<$25 $M_\odot$ yr$^{-1}$ ($L/L^\star$)$^{-1}$; Christensen et al.
2004). We thus conclude that the host galaxy of GRB000911 is a very
compact, subluminous, slightly reddened, extreme starburst galaxy.

Concerning the age of the dominant stellar population, when we make a
comparison with the sample of Christensen et al. (2004) we see that,
according to our best-fit results, the GRB000911 host falls at the high
end of the age distribution. This would suggest that the dominant stellar
population is older than the one of the GRB progenitor; this age implies
that this main population formed at a redshift $z \sim$ 1.3. A similar
but more extreme result was found by Levan et al. (2004) for the case of
the host galaxy of GRB030115. A relatively old age for the GRB000911 host
stellar population is mainly inferred by the presence of a discontinuity
between the $R$ and $I$ optical bands, which is interpreted by {\sl
Hyperz} as the Balmer break placed around 4000 \AA~restframe. We however
caution that it is difficult to distinguish older populations from dusty
environments in high-redshift galaxies with the use of photometric
techniques alone (Moustakas et al. 2004).

The host reddening is indeed moderate but within the range of the GRB host
samples of Sokolov et al. (2001) and Christensen et al. (2004). No signs
of significant intrinsic extinction are however seen in the afterglow SFD;
this is possibly because the dust surrounding the GRB may have been
destroyed by the initial high energy and UV radiation (Waxman \& Draine
2000; Fruchter et al. 2001). This is one of the few hosts for which NIR
data are available (Bloom et al. 1998; Djorgovski et al. 1998; Fruchter et
al. 1999a,b; Frail et al. 2002). Its dereddened $(V-H)_{\rm AB}$
color\footnote{For a given frequency $\nu$, the corresponding AB magnitude
is defined (see, e.g., Bolzonella et al. 2000b) as $m_{\rm AB}$ = $-$2.5
log~$F(\nu)$ $-$ 48.60, where $F(\nu)$ is given in erg s$^{-1}$ cm$^{-2}$
Hz$^{-1}$.}, 1.12$\pm$0.46, compares well with those of other GRB host
galaxies (Fruchter et al. 1999b). Deep imaging of this rather compact host
with HST, however, would allow a more detailed study of its morphology and
other properties.

\section{Conclusions}

The late-epoch and long-term optical/NIR photometric monitoring of the
GRB000911 OT presented here allowed an in-depth analysis of the host
galaxy properties and, in turn, an accurate study of the light curves of
the GRB afterglow. The host SFD indicates that this is a subluminous
extreme starburst compact galaxy with moderate intrinsic extinction, a
luminosity $\sim$0.1$L^\star$, an age of $\sim$0.5 Gyr, a SFR of $\sim$2.7
$M_\odot$ yr$^{-1}$ and a specific SFR of $\approx$30 $M_\odot$ yr$^{-1}$
($L/L^\star$)$^{-1}$. This is the highest specific SFR value for a 
GRB host inferred from optical/NIR data. The GRB000911 afterglow seems to
follow an adiabatic evolution typical of a synchrotron fireball with
electron distribution index $p \sim$ 3 and beaming angle $\theta_0 >$
10$^{\circ}$. The OT light curves show a single power-law temporal decay,
modified at late times by light from the host galaxy, and possibly with a
deviation at late epochs due to an emerging SN contribution. This
component is more apparent in the NIR and could be modeled with a SN
similar, in luminosity and overall shape, to SN 1998bw. In this
case, it would be both the farthest SN connected with a GRB and
the farthest core-collapse SN observed up to now.

\begin{acknowledgements}

We are grateful to the anonymous referee for several useful comments which
helped us to improve the paper. We thank the staff astronomers of the ESO
(La Silla and Paranal), NOT and TNG observatories. We acknowledge Scott
Barthelmy for maintaining the GRB Coordinates Network (GCN) and BACODINE
services. We are grateful to S. Covino for having cross-checked his
$J$-band data calibration with ours and to D. Lazzati for helpful
discussions. We also thank S. Savaglio and E. Rol for useful comments. 
We acknowledge benefits from collaboration within the EU FP5 Research
Training Network ``Gamma-Ray Bursts: An Enigma and a Tool". N. Masetti
acknowledges support under CRUI `Vigoni' programs 31-2002 and 161-2003. 
S. Klose acknowledges support from the DAAD under grants D/0237747 and
D/0103745. This work was partly supported by the Danish Natural Science
Reseach Foundation (SNF) and the Carlsberg Foundation. This research has
made use of the SIMBAD database, operated at CDS, Strasbourg, France and
of NASA's Astrophysics Data System.

\end{acknowledgements}

\end{document}